\documentclass{moriond}

\def\Journal#1#2#3#4{{#1} {\bf #2}, #3 (#4)}

\def\RPP{\em Rep.~Prog.~Phys.}
\def\CQG{\em Class.~Quantum Grav.}
\def\ApJ{\em Astrophys.~J.}
\def\MNRAS{\em Mon.~Not.~R.~Astron.~Soc.}
\def\AA{\em Astron.~Astrophys.}
\def\AAR{\em Astron.~Astrophys.~Rev.}
\def\PRL{\em Phys.~Rev.~Lett.}
\def\PRD{\em Phys. Rev. D}

\def\LLR{\em Living.~Rev.~Rel.}
\def\IJMP{\em Int.~J.~Mod.~Phys.}
\def\SovJ{\em Sov.~J.~Exp.~Theor.~Phys.}
\def\ZhExp{\em Zh.~Eksp.~Theor.~Fiz.~Pisma Red.}
\def\JCAP{\em JCAP}

\def\be{\begin{equation}}
\def\ee{\end{equation}}
\def\bea{\begin{eqnarray}}
\def\eea{\end{eqnarray}}

\newcommand{\Photo}{\includegraphics[height=35mm]{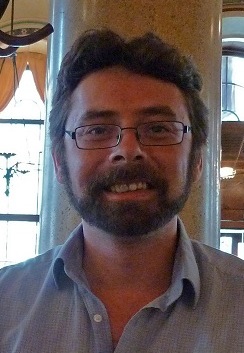}}

\begin{document}
\vspace*{4cm}
\title{GRAVITATIONAL WAVE SEARCHES WITH ADVANCED LIGO AND ADVANCED VIRGO}

\author{
C.~VAN DEN BROECK\\
for the LIGO Scientific Collaboration and the Virgo Collaboration}

\address{
Nikhef -- National Institute for Subatomic Physics,\\
Science Park 105, 1098 XG, Amsterdam, The Netherlands}

\maketitle
\abstracts{
Advanced LIGO and Advanced Virgo are expected to make the first direct detections of gravitational waves (GW) in the next several years. Possible types of GW emission include short-duration bursts, signals from the coalescence of compact binaries consisting of neutron stars or black holes, continuous radiation from fast-spinning neutron stars, and stochastic background radiation of a primordial nature or resulting from the superposition of a large number of individually unresolvable sources. We describe the different approaches that have been developed to search for these different types of signals. In this paper we focus on the GW detection methods themselves; multi-messenger searches as well as further science enabled by detections are dealt with in separate contributions to this volume.}

\section{Introduction}

After periods of commissioning as well as scientific observing runs between 2002 and 2011, the two initial LIGO\,\cite{LIGO} detectors in the US and the Virgo\,\cite{Virgo} observatory in Italy reached their design sensitivities, confirming large-scale laser interferometry as a highly promising technique for the direct detection of gravitational waves (GW). The next phase in these projects consists of the upgrades to Advanced LIGO\,\cite{aLIGO} and Advanced Virgo\,\cite{AdV}, with a gradual improvement in sensitivity of approximately a factor of ten in the course of the next several years, ultimately increasing the volume of space that can be searched by about three orders of magnitude. The Advanced LIGO interferometers will already have a three-month observing run starting in September 2015, to be joined by Advanced Virgo for a six-month run in 2016-17 and a nine-month run in 2017-18, alternated with periods of commissioning; design sensitivity is expected to be reached around 2019.  Exacty when these observatories will have their first detections depends on the instruments' duty cycles as well as astrophysical events rates (the latter being highly uncertain), but a GW observation of binary neutron star coalescence before the end of the decade is very plausible\,\cite{ObservingScenarios}. In a few years' time the KAGRA\,\cite{KAGRA} detector in Japan will join the Advanced LIGO-Virgo network, and LIGO-India\,\cite{LIGOIndia} may become active around 2022. The smaller GEO-HF\,\cite{GEO} in Germany is already taking data.

The expected GW signals can be divided into transient signals, whose duration might be anywhere between a millisecond and several hours, and long-duration signals that are continuously emitted. Among the most promising sources of transient GW are \emph{compact binary coalescences} of two neutron stars, a neutron star and a black hole, or two black holes; in this case theoretical predictions of the waveform shapes are available, and coalescence events can be searched for by comparing these with the data. As we shall see, there are many other possible transient sources of GW, called \emph{bursts}, whose emission is much more difficult to predict in detail; the absence of a waveform model then necessitates different ways of searching. Examples of potentially detectable long-duration signals are \emph{continuous waves} from fast-spinning neutron stars, either isolated or accreting matter from a companion star. In this case the emission happens essentially at a single frequency, which however may be changing due to neutron star spin-down and which will be Doppler-modulated due to orbital motion in a binary as well as the motion of the Earth; when searching for unknown stars this leads to a computationally challenging problem. Finally, there will be searches for \emph{stochastic backgrounds} coming from all directions on the sky; these could be of a primordial origin (\emph{e.g.}~inflation, or phase transitions in the early Universe), or they could arise from the superposition of a large number of point sources that are not individually resolvable. 

Once a GW detection has been made, we will want to characterize the signal and, if possible, reconstruct the source. This involves the development of parameter estimation techniques, which again differ depending on the extent to which the source can be theoretically modeled. We will briefly discuss the methods that are being pursued; what further science this enables is the subject of a separate contribution to these Proceedings.

In this paper we will deal with the detection of the GW signals themselves, but it should be noted that GW searches can be combined with searches for associated electromagnetic and/or neutrino emission; these \emph{multi-messenger} efforts will also be discussed in a separate paper in this volume.

\section{Gravitational wave searches}

We now discuss in turn the searches that will be performed for compact binary coalescences, unmodeled or poorly modeled burst events, continuous waves from fast-spinning neutron stars, and stochastic backgrounds, along with methods for parameter estimation. 

\subsection{Compact binary coalescence}


Coalescing compact binaries consisting of two neutron stars (NS-NS), a neutron star and a black hole (NS-BH), or two black holes (BH-BH), with typical black hole masses up to a few tens of solar masses ($M_\odot$), are among the most promising sources for the advanced detectors, visible out to hundreds of megaparsecs. The emission of gravitational waves leads to loss of orbital angular momentum as well as orbital energy, causing orbits to circularize as well as shrink. If the component objects do not have intrinsic spins then the GW emission in this quasi-circular \emph{inspiral} regime has a gravitational waveform whose frequency and amplitude both increase steadily as a `chirp'; when spins are non-zero then the waveform can get modulated due to precession of the orbital plane caused by spin-orbit and spin-spin interactions\,\cite{Apostolatos}. The inspiral continues until some last stable orbit is reached, after which the components plunge towards each other and \emph{merge}, leading to a single black hole (or possibly a hypermassive neutron star). Such a black hole will be highly excited and undergo \emph{ringdown} as it asymptotes to  a quiescent, Kerr state. The early inspiral is well-understood in terms of post-Newtonian series expansions in $v/c$, with $v$ some characteristic velocity\,\cite{Blanchet}; similarly, the ringdown signal can be described using black hole perturbation theory\,\cite{Kokkotas}. In recent years, large-scale numerical simulations have aided analytical waveform modeling in the construction of phenomenological models that exhibit a high degree of faithfulness with numerical waveforms also in the late inspiral and merger regimes\,\cite{Santamaria,Sturani,Taracchini,Schmidt}.

Thus, the data analyst is increasingly well-equipped with waveform models that can be used to efficiently search for compact binary coalescence signals\,\cite{NINJA-2}. The main method used is \emph{matched filtering}\,\cite{matchedfiltering}, where one integrates the data against trial waveforms divided by a detector's noise power spectral density, to give more weight to frequencies where the instruments are the most sensitive; the resulting number is the signal-to-noise ratio (SNR). SNRs are computed for many different choices of the intrinsic parameters (masses and spins), which together comprise a `template bank'.\cite{templatebank} Candidate detections must have SNRs above a certain threshold in at least two detectors, at consistent parameter values. Since non-stationarities in the noise can mimick GW events, one further performs \emph{e.g.}~a $\chi^2$ test\,\cite{chisq} to check that the build-up of SNR over frequency is consistent with what one would expect from a real signal. Finally, a noise background distribution is constructed by sliding detector outputs in time with respect to each other and looking for coincident triggers, which are then guaranteed not to be GW events. This background allows one to assign a significance to candidate detections\,\cite{Babak}.

Which waveform approximants are deployed depends on the kind of source one is searching for. In the case of binary neutron star coalescence, the signal terminates at high frequencies and mainly the inspiral part of the waveform is accessible, where post-Newtonian theory is valid to good approximation\,\cite{BIOPS}; moreover, neutron stars in binaries are expected to have small instrinsic spins\,\cite{Kramer}, so that at first instance one may choose to neglect them. By contrast, for NS-BH and BH-BH coalescences, all three regimes of inspiral, merger, and ringdown are in the detectors' sensitive frequency band, and astrophysical black holes are likely to have large spins\,\cite{Kalogera}, leading to the abovementioned precession-induced modulation of the signal. Despite significant efforts by the waveform modeling and data analysis communities, template banks that include precession effects are not yet available. Although the use of waveforms with non-zero, aligned spins has been shown to significantly boost the sensitivity of searches compared with non-spinning templates\,\cite{Harry}, some fraction of signals that are in principle detectable will still be missed by not taking precession into account\,\cite{Canton}. Part of the solution may be the use of reduced order modeling, a technique which identifies the essential features of a waveform family and discards the less important information\,\cite{Field,Purrer}.

When performing matched filtering, the parameter space over which templates need to be placed only comprises instrinsic parameters: the masses of the two component objects, and (if needed) their intrinsic spins; for the purpose of detection, other parameters such as the sky location, the orientation of the inspiral plane with respect to the observer, and the distance to the source can be absorbed into an overall amplitude of the waveform. 
A rapid determination of the approximate sky position can be performed by looking at the relative phases, amplitudes, and times of arrival of the signal in the different detectors\,\cite{first2years}. However, a full estimation of the source parameters will require the exploration of the complete parameter space; tools that have been developed to this end include MCMC and nested sampling methods\,\cite{lalinference}.

\subsection{Burst searches}

Aside from binary coalescences there is a host of other transient sources with potentially detectable GW signals. Most of these are not well modeled, so that matched filtering is often not an option; moreover, one will also want to search for transient signals of unknown origin, in which case no signal model can be assumed at all. Burst sources include supernovae\,\cite{Ott}, long gamma ray bursts that may be caused by the gravitational collapse of massive stars\,\cite{Freyer,Metzger}, and soft gamma-ray repeater giant flares in pulsars\,\cite{Mereghetti}. These are sources that can be seen if they occur in our near our galaxy; examples of sources that are in principle observable out to cosmological distances are cusps and kinks on cosmic (super)strings\,\cite{Damour,Olmez}.  Burst signals potentially span a wide range in frequency (a few Hz to several kHz) and time (milliseconds to hours). 

Another target for burst searches are coalescences involving \emph{intermediate mass} black holes with masses up to a few hundred solar masses, which may exist in globular clusters\,\cite{Miller}. Intermediate mass black hole binaries merge at low frequencies ($f < 200$ Hz for total binary mass $M > 50\,M_\odot$), so that the part of the signal in band is dominated by the late inspiral and merger, which are poorly modeled; in that case a burst search is advisable to supplement matched-filter searches\,\cite{IMBH}. Similarly, binaries whose orbits have significant eccentricity are not well understood theoretically, and here too there will be great benefit in performing a burst search\,\cite{Huerta}.

A completely generic burst search requires techniques that can distinguish genuine GW signals from transient noise in the detectors without any prior knowledge of the waveform. In that case one usually first combines data from all detectors in terms of amplitude and phase, such that a GW signal builds up coherently while noise artefacts are removed based on their lack of correlation between detectors. Next the data is decomposed using \emph{e.g.}~short Fourier transforms or wavelets, and candidate signals are identified as `bright' pixels in time-frequency maps\,\cite{cWB1,cWB2,cWB3,Xpipeline,STAMP}. Alternative methods first process the data streams separately (`incoherently') and then look for coincidences between detectors or trigger a coherent MCMC follow-up\,\cite{LIB,Bayeswave}.

Also for burst sources, detection will be followed by parameter estimation. In the case of coalescing binaries, this implies measuring the parameters determining the sources (masses, spins, \ldots). By contrast, for generic burst signals the nature of the source will be \emph{a priori} unknown; what one does in this case is to try and characterize the signal rather than the source, in terms of its time-frequency and polarization content. 

For some burst sources a certain amount of theoretical modeling has been done. For example, in the case of supernovae, large-scale numerical simulations have been performed with different underlying assumptions, \emph{e.g.}~the neutrino mechanism\,\cite{Bethe}, the magnetorotational mechanism\,\cite{Burrows}, or the acoustic mechanism\,\cite{acoustic}. These lead to waveforms that exhibit qualitative differences. Using a principal component decomposition, the main features can be extracted from sets of numerical relativity waveforms resulting from the different assumptions made in the simulations. By performing Bayesian model selection on detected supernova signals, the different supernova models can be ranked, which will give insight into which mechanism dominates\,\cite{Logue}.

At the extreme end, signals from cosmic string cusps and kinks are sufficiently well-understood that template waveforms are available\,\cite{Damour,Olmez}, so that a matched-filter search is in fact possible: in the case of cusps the frequency dependence of the signal is $h \propto f^{-4/3}$, while for kinks one has $h \propto f^{-5/3}$. The amplitudes depend on the string tension, and the intrinsic rate of cusp and kink events depends on the loop size, the string tension, and the reconnection probability, which in the case of superstrings is smaller than one. The non-detection of cosmic string signals with initial LIGO and Virgo has already allowed exclusion of a significant part of parameter space\,\cite{cosmicstringsearch}.

\subsection{Continuous waves from fast-spinning neutron stars}

Fast-spinning neutron stars can be sources of detectable gravitational waves also when not part of a compact binary. GW emission can result from asymmetries due to elastic deformations of the crust\,\cite{Bildsten,Ushomirsky}, deformations through magnetic fields\,\cite{Cutler}, GW-driven unstable oscillation modes (r-modes\,\cite{Andersson}, and f-mode Chandrasekhar-Friedman-Schutz instabilities\,\cite{Friedman}), or free precession arising from a misalignment of a neutron star's symmetry axis and the rotation axis\,\cite{Jones}. In cases where the rotation frequency $f_{\rm rot}$ can be established through electromagnetic observations, comparison with the main gravitational wave frequency $f_{\rm GW}$ can reveal the emission mechanism. If $f_{\rm GW} = 2 f_{\rm rot}$, with no GW emission observed at $f_{\rm rot}$, then the gravitational radiation is mainly due to non-axisymmetric deformation; on the other hand, if there is appreciable GW emission also at $\simeq f_{\rm rot}$ then precession probably plays a role. If $f_{\rm GW} \simeq (4/3)\,f_{\rm rot}$ then r-modes are strongly favored, yielding direct information on the interior fluid motion. 

In the `standard' scenario of non-axisymmetric deformation ($f_{\rm GW}  = 2 f_{\rm rot}$), the amplitude of the GW emission is proportional to the equatorial ellipticity $\epsilon = (I_{xx}-I_{yy})/I_{zz}$, where the $I_{ii}$ are the moments of inertia of the star, with the spin axis in the $z$-direction. Estimates of $\epsilon$ are highly uncertain\,\cite{Ushomirsky}, but might be as large as $10^{-6}$. Neutron stars in orbit with an ordinary star should be able to maintain axisymmetry due to accretion, and the balancing of accretion torque and GW emission may explain why the spin frequencies of known accreting neutron stars never approach the break-up limit\,\cite{Bildsten}. Depending on the typical size of $\epsilon$, the currently known neutron stars may not yield detectable GW signals, although astrophysically interesting upper bounds have already been put on the percentage of the spin-down energy loss that is due to gravitational radiation\,\cite{spindownupperlimits}. However, the galaxy contains an estimated $10^9$ neutron stars, some fraction of which may have eluded electromagnetic detection, yet may be sufficiently close, non-axisymmetric, and fast-spinning to allow for detection with Advanced LIGO and Advanced Virgo.

Though fast-spinning, non-axisymmetric neutron stars are intrinsically quasi-monochromatic, with a frequency that is slowly decreasing, the observed GW signal gets Doppler-modulated due to the motion of the Earth, and by the orbital motion in the case of a binary. GW searches are divided into three categories. \emph{Targeted searches} are those where the neutron star is visible as a pulsar so that its spin frequency (and where applicable, the orbital Doppler frequency) is known from electromagnetic observations\,\cite{TDBayesian,Fstatistic,FGstatistic,5vector}. In \emph{directed searches}, the sky position is known but not the source frequency\,\cite{Fstatistic,Sideband,TwoSpect,CrossCorr,Polynomial}. Finally, in \emph{all-sky searches} one also looks for neutron stars that have not been discovered by electromagnetic means\,\cite{FGstatistic,TwoSpect,Polynomial,SkyHough,FreqHough,PowerFlux}.

In targeted and directed searches, at least some known parameters can be folded into the analysis; in particular, the Doppler modulations can be removed and the search problem reduces to looking for sinusoidal signals; as in the case of compact binary coalescence, intrinsic parameters such as the GW polarization, a reference phase, and the inclination (in the case of binary systems) can be marginalized over. One can then integrate the data over a very long time, in principle the data taking time. By far the most computationally challenging are the all-sky searches. In that case one has to search over all sky positions, frequencies, spin-down parameters, and, in the case of binaries, the orbital parameters. Integration over long periods of time $T$ then becomes difficult, since the resolution in parameter space increases rapidly ($\propto T^5$ even for searches including only leading-order spin-down). In the latter case a hierarchical search must be resorted to in order to keep the problem computationally tractable: one starts with a lower-resolution search that identifies interesting candidates, after which the search is iteratively refined as smaller and smaller parts of parameter space can be studied with increasingly higher resolution.

\subsection{Stochastic searches}

Omni-directional gravitational wave background radiation could arise from fundamental processes in the early Universe, or from the superposition of a large number of signals with a point-like origin. Examples of the former include parametric amplification of gravitational vacuum fluctuations during the inflationary era\,\cite{Grishchuk,Starobinsky}, termination of inflation through axion decay\,\cite{Barnaby} or resonant preheating\,\cite{Easther},  Pre-Big Bang models inspired by string theory\,\cite{Mandic}, and phase transitions in the early Universe\,\cite{Starobinsky}; the observation of a primordial background would give access to energy scales of $10^9 - 10^{10}$ GeV, well beyond the reach of particle accelerators on Earth. Astrophysical confusion backgrounds could result from the collective emission of spinning neutron stars in the Galaxy\,\cite{Regimbau}, compact binary coalescences\,\cite{Wu} out to redshifts of $z \sim 5$, or superpositions of cosmic string bursts\,\cite{Damour,Olmez}. 

Stochastic backgrounds are conveniently described in terms of the GW energy density $\rho_{\rm GW}$ per logarithmic frequency bin, normalized to the critical density of the Universe, $\rho_c$: 
\begin{equation}
\Omega_{\rm GW} = \frac{1}{\rho_c} \frac{d\rho_{\rm GW}}{d \ln f}.
\end{equation}
The current bound from the initial LIGO-Virgo era is\,\cite{improvedupperlimits} $\Omega_{\rm GW} < 6.1 \times 10^{-6}$. Advanced detectors, both through their improved strain sensitivity and their wider frequency sensitivity band, will probe as low as $\Omega_{\rm GW} \sim 6 \times 10^{-10}$. 

Searches for stochastic backgrounds are performed by cross-correlating the output of multiple detectors with an optimal filter that is proportional to an expected form for $\Omega_{\rm GW}$ as a function of frequency\,\cite{Christensen,Allen}. It is reasonable to assume that in the relevant frequency band, stochastic signals can be approximated by a power law: $\Omega_{\rm GW}(f) = \Omega_0 f^\alpha$. The index $\alpha$ will take on different values depending on the origin of the radiation; in blind searches for cosmological sources one tends to quote upper limits on $\Omega_{\rm GW}$ under the assumption of a flat spectrum ($\alpha = 0$), while a superposition of binary inspiral signals corresponds to $\alpha = 2/3$. Although at first instance one may expect stochastic backgrounds  to be largely isotropic, we note that non-isotropic backgrounds, arising from \emph{e.g.}~random fluctuations in the number of point sources, will be searched for as well\,\cite{stochdirected}.

Like in searches for continuous waves from fast-spinning neutron stars, one can integrate over long times, in this case the length of the data set. In terms of backgrounds due to compact binary coalescences, it turns out that a signal can be seen after one year of operation at design sensitivity, and assuming a rate of a few tens of binary neutron star coalescences per year within a distance of a few hundred megaparsecs\,\cite{Wu}. With the latter rate, the dedicated searches for compact binary coalescences are themselves bound to make detections, but since the stochastic background includes sources out to redshifts of several, its measurement can give valuable information about the evolution of star formation rates. 

Finally, one can search for stochastic backgrounds with non-standard polarizations, such as longitudinal modes, `breathing' modes, and vector modes, which are predicted by various alternative theories of gravity\,\cite{Will}. This can be done with a relatively straightforward extension of the existing methods, taking into account the different coherence structure of signals across detectors.

\section{Conclusions}

There exists a wide variety of sources whose GW emission can potentially be detected with second-generation interferometric observatories, within our galaxy (\emph{e.g.}~fast-spinning neutron stars and a range of burst sources), at distances of hundreds of megaparsecs (such as coalescing compact binaries), and at cosmological scales (\emph{e.g.}~primordial GW backgrounds), promising rich scientific returns. For all these, tailored and robust data analysis techniques are in place, which are continually being improved even further. With the construction of Advanced LIGO and Advanced Virgo well underway, we can look forward to the first direct detections of gravitational waves in the next several years.

\section*{Acknowledgments}

The authors gratefully acknowledge the support of the United States National Science Foundation (NSF) for the construction and operation of the LIGO Laboratory, the Science and Technology Facilities Council (STFC) of the United Kingdom, the Max-Planck-Society (MPS), and the State of Niedersachsen/Germany for support of the construction and operation of the GEO600 detector, the Italian Istituto Nazionale di Fisica Nucleare (INFN) and the French Centre National de la Recherche Scientifique (CNRS) for the construction and operation of the Virgo detector. The authors also gratefully acknowledge research support from these agencies as well as by the Australian Research Council, the International Science Linkages program of the Commonwealth of Australia, the Council of Scientific and Industrial Research of India, Department of Science and Technology, India, Science \& Engineering Research Board (SERB), India, Ministry of Human Resource Development, India, the Spanish Ministerio de Econom\'{i}a y Competitividad, the Conselleria d’Economia i Competitivitat and Conselleria d’Educaci, Cultura i Universitats of the Govern de les Illes Balears, the Foundation for Fundamental Research on Matter supported by the Netherlands Organisation for Scientific Research, the Polish Ministry of Science and Higher Education, the FOCUS Programme of Foundation for Polish Science, the European Union, the Royal Society, the Scottish Funding Council, the Scottish Universities Physics Alliance, the National Aeronautics and Space Administration, the Hungarian Scientific Research Fund (OTKA), the Lyon Institute of Origins (LIO), the National Research Foundation of Korea, Industry Canada and the Province of Ontario through the Ministry of Economic Development and Innovation, the National Science and Engineering Research Council Canada, the Brazilian Ministry of Science, Technology, and Innovation, the Carnegie Trust, the Leverhulme Trust, the David and Lucile Packard Foundation, the Research Corporation, and the Alfred P. Sloan Foundation. The authors gratefully acknowledge the support of the NSF, STFC, MPS, INFN, CNRS and the State of Niedersachsen/Germany for provision of computational resources. This research has made use of the SIMBAD database, operated at CDS, Strasbourg, France.

\end{document}